# Surf_2_Volume: A Python Toolbox for Cortical Ribbon Based CIFTI to NIfTI Conversion

*Methodology*

Shuguang Yang[1,2,3]†, Ziyi Wang[1,2,3]†, Yujing Shen[1,2,3], Zhengye Wang[1,2,3], Junyi Li[1,2,3], Yujing Nie[1,2,3], Feizhen Cao[1,2,3]*, Suiping Wang[1,2]*

[1] Philosophy and Social Science Laboratory of Reading and Development in Children and Adolescents, South China Normal University, Ministry of Education, Guangzhou 510631, China

[2] School of Psychology, South China Normal University, Guangzhou 510631, China

[3] Guangdong Key Laboratory of Mental Health and Cognitive Science, South China Normal University, Guangzhou 510631, China

† These authors contributed equally.

* Corresponding authors:

Suiping Wang (wangsuiping@m.scnu.edu.cn); Feizhen Cao (caofeizhen970411@gmail.com).

# Abstract

Many neuroimaging workflows require NIfTI label volumes, whereas cortical parcellations are increasingly distributed in CIFTI surface space. Existing conversion tools offer several mapping settings. In nearest vertex mapping, a small distance can leave cortical voxels unlabeled, whereas a larger distance can extend labels beyond the intended cortical region. Choosing among settings is difficult when no paired volume atlas is available. We present Surf_2_Volume, a Python toolbox that projects discrete cortical labels within an adjustable cortical ribbon. Users can specify the ribbon threshold directly or use an optional procedure that ranks candidate thresholds using information from the source atlas. This procedure ranks 12 thresholds by differences between each label's proportion of source cortical vertices and its proportion of output voxels, without using a published volume atlas for selection. We evaluated four configurations from two atlas families: Yeo2011 parcellations with 7 and 17 networks, represented by 51 and 114 spatial components, and Schaefer2018 100-parcel atlases with 7 and 17 network assignments. All four selected sets included threshold 0.05, which had the highest macro-Dice and macro-Jaccard among the tested package thresholds. Eleven of the 12 selected outputs had higher macro-Dice than the best tested Connectome Workbench mapping for their atlas. In 24 HCP participants, all 12 selected outputs had higher participant-level mean parcel time-series concordance than every tested Workbench mapping (all within-atlas FDR-adjusted $p < 0.001$). Surf_2_Volume provides a reproducible way to generate and select volume atlas candidates when only a surface parcellation is available, while retaining alternatives for different coverage and signal-extraction needs.

# 1. Introduction

The Human Connectome Project (HCP) provides large, consistently processed datasets for studying brain structure, function, and connectivity (Glasser, Smith, et al., 2016). Several widely used parcellations derived from HCP processing use the CIFTI format. CIFTI represents the cerebral cortex on surface vertices and subcortical structures as volume voxels (Glasser, Coalson, et al., 2016; Ji et al., 2019). This representation follows cortical geometry and avoids unnecessary sampling of white matter and cerebrospinal fluid.

However, many neuroimaging programs work with NIfTI volumes and cannot directly use CIFTI label files. This includes workflows based on SPM (Ashburner, 2012) and FSL (Jenkinson et al., 2012) and visualization with MRIcroGL (Rorden, 2025). Published projections are available for some atlases, including HCP-MMP1.0 on fsaverage (Mills, 2016) and an extended HCP multimodal atlas in MNI152 space (Huang et al., 2022), but they do not provide a general method for converting other CIFTI parcellations.

Connectome Workbench (Marcus et al., 2011) can map surface labels to a volume using either cortical ribbon geometry or the nearest vertex. In nearest vertex mapping, the maximum distance determines how far from the surface a voxel can receive a label. A small distance restricts labeling to voxels close to the surface and can leave much of the cortex unlabeled. Increasing the distance extends the same nearest vertex assignment to more distant voxels, improving coverage but potentially expanding labeling beyond the cortex. The representative outputs in Fig. 2 illustrate the contrast between limited coverage and broad labeling. This makes parameter choice consequential: the setting that labels the most voxels may differ from the setting that best reproduces parcel boundaries or parcel signals.

A published volume version of an atlas provides a reference against which candidate settings can be compared. When such a counterpart is unavailable, overlap and coverage against that reference cannot guide parameter selection. Researchers therefore need a way to control where labels are assigned and to choose plausible settings using the surface atlas itself. These are related tasks: defining the permitted labeling region limits spatial expansion, while a criterion derived from the input atlas gives users a reproducible basis for comparing candidate conversions.

We developed Surf_2_Volume to perform this conversion through a Python function interface. The package assigns discrete labels within a target cortical ribbon, the volume between the inner and outer cortical surfaces, and uses a threshold to adjust the region eligible for filling. Users can specify this threshold manually or let the package compare candidates using the proportions of cortical vertices and output voxels assigned to each label. The automatic procedure does not require a published NIfTI version of the input atlas. Its conversion workflow does not invoke external neuroimaging executables. The package bundles reference assets for MNI152NLin6Asym (Evans et al., 2012) and supports other MNI152 standard spaces when compatible reference files are provided.

We evaluated Surf_2_Volume using the Yeo2011 parcellations with 7 and 17 networks (Thomas Yeo et al., 2011), followed by the Schaefer2018 100-parcel atlases with 7 and 17 network assignments (Schaefer et al., 2018). The Yeo configurations were represented by 51 and 114 spatially connected components. This design tests candidate selection across two atlas families with different spatial organization. After selection, converted volumes were compared with their paired volume references and with Connectome Workbench (WB) mappings using spatial overlap and coverage measures. HCP preprocessed resting-state data provided a downstream test of agreement between parcel time series extracted from each conversion and its paired volume reference.

# 2. Materials and Methods

## 2.1 Atlas and paired-reference design

Spatial evaluation used four atlas configurations from two families. We first evaluated the Yeo et al. (2011) 7- and 17-network parcellations derived from the 1,000-participant reference sample (N1000), using the SplitLabels representations distributed by the authors through CBIG. In these representations, disconnected regions of a network receive separate identifiers, giving 51 and 114 spatial components, respectively. The corresponding fsLR-32k CIFTI label maps served as surface inputs, and the paired component-labeled FSL MNI152 1 mm volumes served as evaluation references. We then evaluated the Schaefer et al. (2018) 100-parcel atlases with 7 and 17 network assignments and their author-released volume references. The two Schaefer configurations share the 100-parcel spatial definitions and differ in network assignment. We refer to the configurations as Yeo7, Yeo17, Schaefer7, and Schaefer17. Each prediction was evaluated on its paired reference grid. All four benchmarks were cortical; the package's optional retention of subcortical labels was not tested here.

## 2.2 Processing steps in the Python package Surf_2_Volume

Surf_2_Volume can be installed from TestPyPI with *pip install -i https://test.pypi.org/simple/ surf2volume==0.2.0*. The entry function, *cifti_to_volume()*, coordinates label separation, surface transfer, projection into the target ribbon, and volume merging (Fig. 1). It uses NiBabel (Brett et al., 2020), neuromaps (Markello et al., 2022), NumPy (Harris et al., 2020), and SciPy (Virtanen et al., 2020). The conversion separates two operations: the target ribbon defines where assignment is permitted, and categorical voting determines which label an eligible voxel receives. Adjusting the ribbon threshold therefore changes the permitted spatial extent without interpolating between label identifiers.

Required geometric inputs are the source and target registration spheres, matching target smoothwm and pial surfaces, target ribbon volumes, and mesh-to-mesh (M2M) correspondence files. The smoothwm surface approximates the inner cortical boundary and the pial surface the outer boundary. Bundled references are provided for MNI152NLin6Asym (Evans et al., 2012). Other target spaces require compatible files and the corresponding *subjects_dir* and *target_subject* settings. Surface coordinates and ribbon image coordinates must describe the same target space; the ribbon affine converts surface coordinates to voxel indices.

First, *separate_cifti()* reads the CIFTI label and brain-model axes and separates left and right cortical labels. Each cortical array is placed at the vertex indices specified by its brain model, with unassigned vertices set to background. Any subcortical labels are extracted into a volume using the CIFTI volume geometry. Each hemisphere is then processed against its matching reference surfaces, preserving the input label identifiers.

Neuromaps resamples cortical labels within fsLR and then to fsaverage, using the nearest option for label data. *resample_surface_label()* transfers these labels to the target surface by finding the nearest source vertex on the registered spheres. *resample_gifti_with_m2m()* then assigns labels to the target projection mesh using the M2M file's NearestNode correspondences. These transfers operate on discrete labels and retain their identifiers. Surface transfers are performed once per input atlas and reused across the threshold candidates.

Next, *spackle_surface_to_volume()* smooths each hemisphere's ribbon image at 2 mm full width at half maximum (FWHM). Voxels whose smoothed ribbon values lie between the candidate lower threshold and the fixed upper bound of 2.0 are eligible for labeling. A lower threshold admits more voxels near the ribbon boundary, whereas a higher threshold restricts this support. Smoothing applies to the ribbon image, not to the parcel labels. The same smoothed ribbon is reused for all candidate thresholds.

For initial projection, corresponding smoothwm and pial vertices define a sampling path through the cortical thickness. The default uses 10 equally spaced points from 10% of the smoothwm-to-pial distance before the inner surface to 10% beyond the outer surface. Sample coordinates are converted to the nearest voxel index using the inverse ribbon affine. Assignment proceeds through the first continuous segment of eligible samples along each path and stops when that segment leaves the mask. Repeated visits by the same vertex to the same voxel contribute only one vote. A voxel receives the most frequent nonzero label among the vertices that reach it; ties are resolved by the smallest label identifier.

Unlabeled voxels that remain inside the eligible mask are then filled from labeled neighbors within a spherical radius of 2 mm. Each iteration assigns the modal nonzero label in that neighborhood, with ties again resolved by the smallest identifier. Filling uses the labels present at the start of the iteration and is repeated for at most four iterations. The process stops earlier when no holes remain or no further voxel can be filled. Voxels outside the mask remain background. This local filling step improves continuity within the permitted region while retaining categorical labels.

Finally, *merge_label_volumes()* combines the left hemisphere, right hemisphere, and optional subcortical volume in that priority order, with later inputs filling only zero-valued voxels. Inputs on different grids are first resampled to the left hemisphere grid using nearest-neighbor assignment. The output is an integer NIfTI label volume with that grid's affine. Intermediate files are retained for inspection. Before evaluation, we checked image shape, affine, positive label identifiers, and negative voxels after alignment to the paired reference grid.

Bundled fsaverage registration spheres came from FreeSurfer 7.4.1 (Fischl, 2012). Target MNI152NLin6Asym surfaces and ribbon references were reconstructed locally with FreeSurfer 7.4.1; SUMA surfaces and M2M files were prepared with AFNI/SUMA (Saad & Reynolds, 2012), as recorded in the package manifest. These prepared assets are used during conversion. Label transfer, ribbon projection, filling, and threshold scoring are performed through the package's Python workflow without invoking external neuroimaging executables.

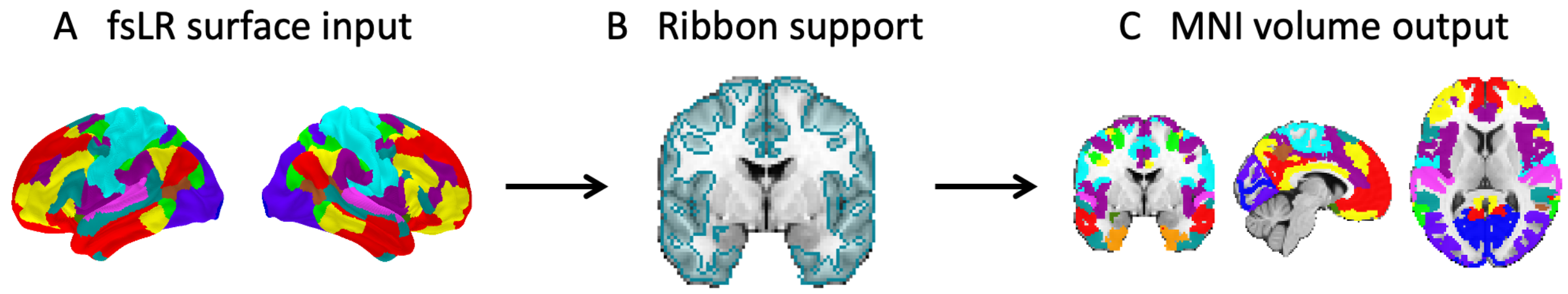

**Fig. 1** Surface-to-volume conversion with Surf_2_Volume. A: categorical cortical labels in fsLR surface space. B: target cortical-ribbon support used for threshold-based filling. C: categorical NIfTI volume output in MNI space. The package performs surface resampling, target-mesh mapping, ribbon filling, and merging of hemispheric outputs; subcortical labels can also be retained when present in the input.

### 2.3 Automatic threshold selection and mapping comparators

With *thresholds="auto"*, *cifti_to_volume()* generates a volume for each candidate lower ribbon threshold while keeping the geometric inputs and other mapping parameters fixed. The tested thresholds were 0.01, 0.025, 0.05, 0.075, 0.10, 0.15, 0.20, 0.25, 0.30, 0.40, 0.50, and 0.75. *select_best_threshold()* scores each candidate by comparing the relative representation of cortical labels in the source CIFTI and the output volume. This gives every candidate an explicit score derived from the input atlas, without requiring a paired published volume atlas.

Let K be the number of valid cortical label identifiers observed in the source left and right cortical brain models, excluding background (label 0) and unlabeled entries. For label k, let $p_k$ be its source cortical vertex count divided by the total vertex count over these K labels, and $q_k(t)$ its output voxel count at threshold t divided by the total output voxel count over the same labels. The discrepancy score is:

$$D(t) = \frac{1}{K}\sum_{k=1}^{K} |p_k - q_k(t)| \quad (1)$$

A smaller D(t) indicates closer agreement in relative label proportions. An output missing a source label assigns that label a count of zero. A candidate with no matching cortical labels receives an infinite score and is excluded. Candidates are ranked in ascending score order, with ties retaining the input threshold order. In automatic mode, the current implementation returns the three highest-ranked thresholds and their volumes, writes candidate scores and selected ranks to auto_threshold_selection.json, and produces a comparison plot. Thresholds reported below in numerical order identify set membership rather than score rank.

Selection uses only the source CIFTI and generated candidate volumes. The paired volume references and HCP time-series concordance do not enter the score or ranking and are used only for subsequent evaluation. The score measures relative label proportions; it does not directly measure local boundary agreement. The target geometric and ribbon assets described in Section 2.2 are required for conversion regardless of whether a paired atlas volume is available.

WB comparators were generated with *wb_command -label-to-volume-mapping* after separating CIFTI cortical labels into left and right GIFTI files. The mapping script used S1200 fsLR-32k midthickness surfaces for mapping coordinates, with matching white and pial surfaces for ribbon mapping. Nearest vertex maximum distances were 1, 2, 3, 5, and 7 mm. Ribbon mapping used the default voxel subdivision of 3, without the optional greedy or thick-column settings. The volume-space input supplied the target image geometry. Positive left labels were written first, with right labels filling remaining background. All outputs were aligned to the paired FSL MNI152 1 mm reference grids by nearest-neighbor resampling before spatial evaluation. HCP time-series comparisons used WB ribbon and nearest 1, 3, 5, and 7 mm; nearest 2 mm was assessed only spatially.

## 2.4 Spatial metrics

All primary spatial evaluations compared each predicted NIfTI with its paired volume reference. Dice and Jaccard were calculated separately for each positive label and averaged without weighting: 51 components for Yeo7, 114 components for Yeo17, and 100 parcels for either Schaefer configuration. Macro-Dice was the primary overlap measure because of its widespread use in neuroimaging segmentation and atlas comparison (Müller et al., 2022; T. Eelbode et al., 2020; Taha & Hanbury, 2015); macro-Jaccard was a stricter complementary measure. For Yeo, the macro-average is therefore over spatial components rather than the original network identifiers.

The missing reference fraction was the number of voxels labeled in the reference but zero in the prediction, divided by the reference support. The extra-reference fraction was the number of voxels labeled in the prediction but zero in the reference, divided by the predicted support. Voxel-wise label accuracy was the fraction of reference-support voxels for which the predicted parcel identifier exactly matched the reference; missing voxels therefore counted as errors.

## 2.5 HCP parcel time-series agreement

We selected preprocessed 3T resting-state data using the HCP-Young Adult 2025 (Glasser, Smith, et al., 2016) download interface, with Subject Group: 100 Unrelated Subjects and Quarter Released: S500. Restricting selection to one release quarter was intended to improve consistency across participants; these filters yielded 24 participants, all included for each of the four atlas configurations. We used the supplied rfMRI_REST aggregate, which combines multiple resting-state runs, and did not preprocess the raw images. The resulting series contained approximately 4,800 time points per participant. Each NIfTI atlas was aligned to the participant's functional grid by nearest-neighbor resampling. For each parcel or Yeo component, the BOLD series was the voxel-wise mean after excluding voxels with a non-finite value at any time point. Series were extracted from the three selected Surf_2_Volume volumes, the five WB volumes specified above, and the paired volume reference. Lin's concordance correlation coefficient (CCC; Lin, 1989) compared each candidate series with its corresponding reference series. CCC values across the 51 Yeo7 components, 114 Yeo17 components, or 100 Schaefer parcels were averaged within each participant, yielding one mean CCC per participant, atlas, and method.

CCC was calculated as twice the covariance of the candidate and reference series divided by the sum of their variances and their squared mean difference. It therefore assesses agreement in signal level and scale as well as covariation. Empty parcels or parcels without any voxel finite at every time point were treated as extraction errors. A zero CCC denominator gives an undefined value; constant series are not otherwise excluded solely for being constant. All evaluated series yielded finite CCC values for every label, participant, and candidate method in each configuration, so participant means used the same complete label set across methods.

### 2.6 Statistical analysis

Spatial metrics were descriptive because each setting yielded one atlas volume per configuration. For time-series inference, we compared participant-level mean CCCs in 15 planned paired t tests per atlas (three Surf_2_Volume thresholds × five WB mappings). Benjamini–Hochberg false-discovery-rate (FDR) correction (Benjamini & Hochberg, 1995) was applied separately to the 15 p values within each of the four atlas configurations. Mean paired CCC differences and their 95% confidence intervals were reported for comparisons with WB nearest 5 mm. Participants were the inferential unit; parcels or components were averaged within participants and were not treated as independent observations.

## 3. Results

### 3.1 Grid and label quality control

After grid harmonization, all predictions for the four configurations matched their paired references in image shape and affine on a 182 × 218 × 182 grid with 1 mm voxels. Yeo7 and Yeo17 outputs retained all 51 and 114 component identifiers, respectively, and each Schaefer output retained all 100 parcel identifiers. No extra or missing label identifiers or negative voxels were detected. Differences in the spatial scores therefore describe label placement and extent rather than loss of entire labels or mismatched image dimensions.

### 3.2 Qualitative comparison of conversion outputs

The three methods produced visibly different coverage patterns in the representative Schaefer7 slices (Fig. 2). At threshold 0.05, Surf_2_Volume labels follow a largely continuous cortical ribbon (Fig. 2A). WB nearest 7 mm labels a broader region, extending beyond the apparent ribbon in parts of the displayed anatomy (Fig. 2B). WB ribbon, default leaves substantial portions of the paired volume reference unlabeled (Fig. 2C).

These slices show how the choice of mapping support affects the converted atlas. Surf_2_Volume provides relatively continuous labeling within a region controlled by the target ribbon threshold. Fig. 2 illustrates representative outputs; the change in coverage and expansion across all tested nearest vertex distances is quantified in Fig. 3 and Tables 1–4.

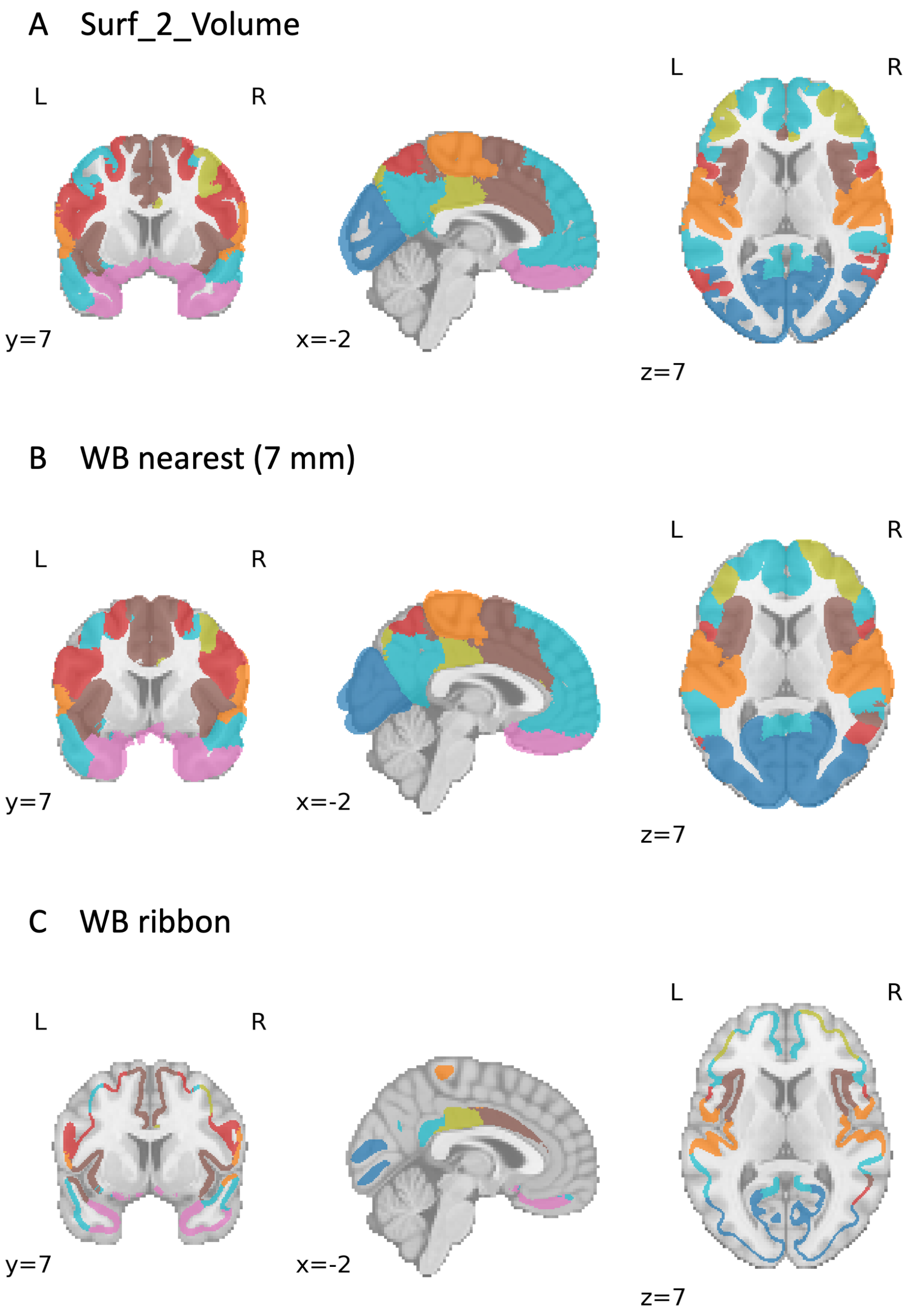

**Fig. 2** Representative volumetric conversion outputs for the Schaefer7 atlas in coronal (y = 7), sagittal (x = −2), and axial (z = 7) views. A: Surf_2_Volume at threshold 0.05. B: WB nearest 7 mm. C: WB ribbon, default. L and R indicate the left and right hemispheres. WB denotes Connectome Workbench. Quantitative results for all four atlas configurations are reported in Fig. 3 and Tables 1–4.

### 3.3 Automatic threshold selection and spatial comparison across four atlases

Automatic selection returned thresholds 0.025, 0.05, and 0.075 for Yeo7, and 0.01, 0.025, and 0.05 for Yeo17 and both Schaefer configurations, without consulting their paired volume references. In all four configurations, the selected set included threshold 0.05, which achieved the highest macro-Dice and macro-Jaccard among the 12 tested package thresholds (4/4 selected sets). Across the three outputs retained per configuration, 11 of 12 had higher macro-Dice than the best tested WB setting for the same atlas. The exception was Yeo17 at 0.01 (0.513 versus 0.520 for WB nearest 5 mm). All 12 outputs also had higher mean HCP reference-series CCC than every tested WB mapping, as detailed in Section 3.4. These counts summarize candidate outputs across the four configurations.

The WB nearest distance sweep showed why broader coverage alone did not identify the best setting. In Yeo7, increasing the maximum distance from 1 to 5 to 7 mm reduced missing reference voxels from 84.63% to 19.60% to 7.13%, while extra-reference labeling increased from 23.95% to 32.65% to 40.41% (Table 1). The highest WB macro-Dice occurred at 5 mm (0.573), rather than 7 mm (0.558). Thus, the largest tested distance covered more of the reference but reproduced its labeled regions less closely.

For Yeo7, selected thresholds 0.025, 0.05, and 0.075 yielded macro-Dice values of 0.590, 0.598, and 0.597 and macro-Jaccard values of 0.435, 0.443, and 0.443 (Fig. 3A; Table 1). Each selected setting exceeded WB nearest 5 mm, the strongest WB overlap comparator (Dice 0.573; Jaccard 0.420). At threshold 0.05, missing support was 12.26%, compared with 19.60% for WB nearest 5 mm; extra-reference labeling was 31.96% versus 32.65%, and voxel label accuracy was 82.03% versus 76.95%. The selected outputs combined higher overlap with closer reference-support agreement than the best-overlapping WB setting.

For Yeo17, thresholds 0.01, 0.025, and 0.05 yielded macro-Dice values of 0.513, 0.525, and 0.533 and macro-Jaccard values of 0.363, 0.374, and 0.381 (Fig. 3B; Table 2). Thresholds 0.025 and 0.05 exceeded WB nearest 5 mm (Dice 0.520; Jaccard 0.368), whereas 0.01 had slightly lower macro-overlap. At 0.05, missing support was 11.80% versus 19.43%, extra-reference labeling was 41.65% versus 42.42%, and voxel label accuracy was 79.33% versus 75.30% for WB nearest 5 mm. The automatically selected set therefore contains useful alternatives rather than three settings that necessarily optimize every reference-based metric.

For Schaefer100, macro-Dice increased from 0.720 at threshold 0.01 to 0.728 at 0.025 and 0.730 at 0.05 in the 7-network atlas, and from 0.719 to 0.727 and 0.729 in the 17-network atlas. Macro-Jaccard increased from 0.567 to 0.581 and from 0.566 to 0.580, respectively (Fig. 3C, D; Tables 3 and 4). All three selected thresholds exceeded the highest WB macro-overlap. Increasing the threshold from 0.01 to 0.05 increased missing support from 5.09% to 11.60%, reduced extra-reference labeling from 16.66% to 9.59%, and reduced voxel label accuracy from 78.38% to 73.57% in Schaefer7 and from 78.35% to 73.54% in Schaefer17.

Among the WB Schaefer mappings, nearest 5 mm had the highest macro-overlap (Dice 0.709; Jaccard 0.556). Compared with that setting, Surf_2_Volume at 0.05 improved macro-Dice by 0.021 and 0.020 in Schaefer7 and Schaefer17, and macro-Jaccard by 0.025 and 0.024. Missing support was lower (11.60% versus 20.57%), as was extra-reference labeling (9.59% versus 11.22%); voxel label accuracy was higher (73.57%/73.54% versus 68.94%). WB ribbon missed 73.83% of reference support and had macro-Dice 0.358, consistent with the incomplete coverage in Fig. 2C. Increasing the WB nearest distance from 1 to 5 to 7 mm reduced missing support from 84.93% to 20.57% to 7.94%, while extra-reference labeling increased from 0.65% to 11.22% to 21.08% in Schaefer7 (Table 3).

Across all four configurations, threshold 0.05 exceeded every tested WB mapping in macro-Dice and macro-Jaccard and improved missing support, extra-reference labeling, and voxel label accuracy relative to WB nearest 5 mm. WB nearest 7 mm retained lower missing support and higher voxel label accuracy, but had larger extra-reference fractions and lower macro-overlap. Yeo extra-reference fractions were larger than Schaefer fractions, reflecting different atlas support. This metric denotes labels in reference-background voxels and does not identify their tissue type. The broader validation supports a consistent overlap advantage while retaining atlas-specific coverage trade-offs.

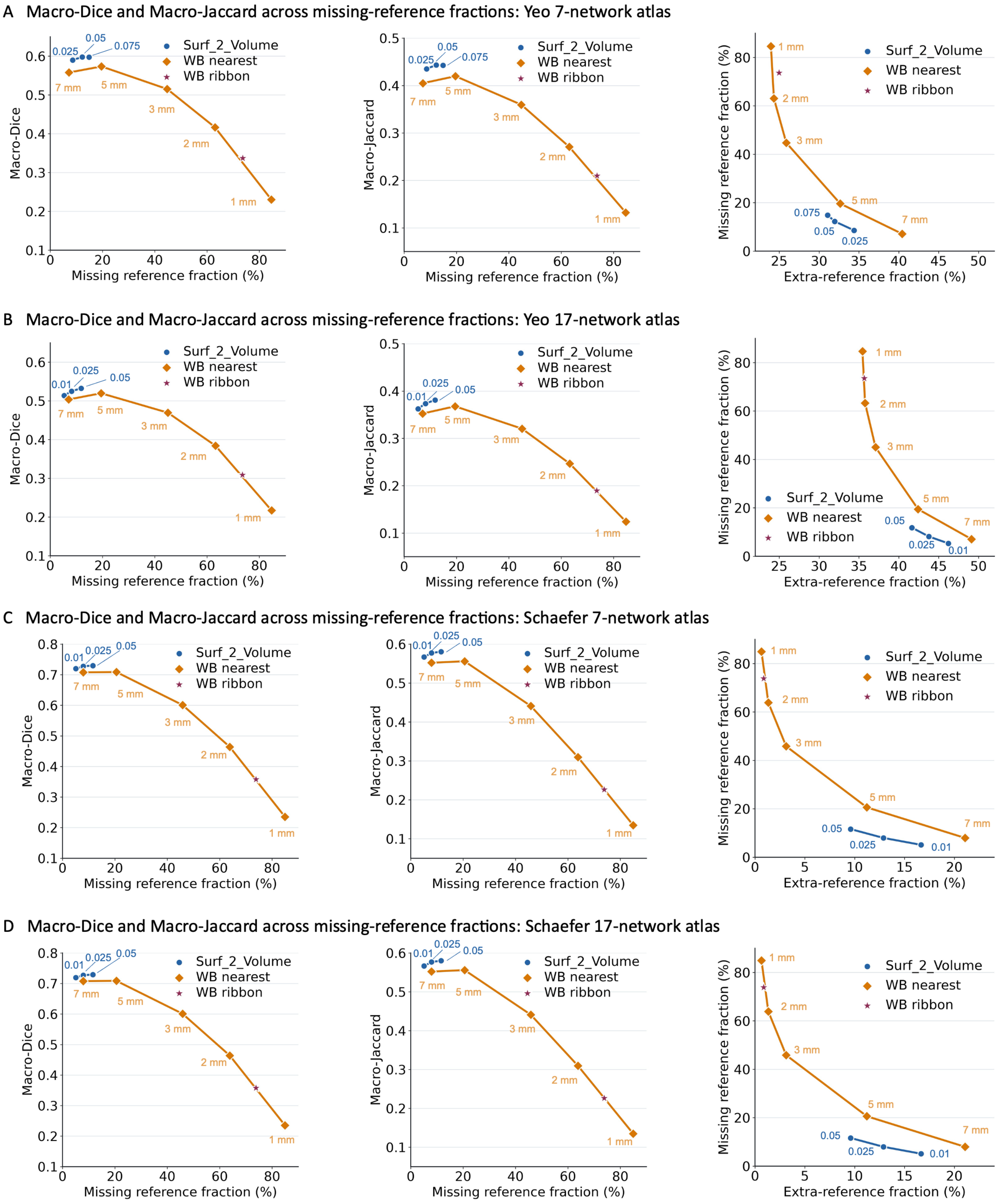

**Fig. 3** Spatial overlap and coverage across four atlas configurations. Rows A–D show Yeo7, Yeo17, Schaefer7, and Schaefer17, respectively. The left and middle plots show macro-Dice and macro-Jaccard against missing-reference fraction; the right plots show missing-reference against extra-reference fraction. Support fractions are percentages. Blue circles denote the three automatically selected Surf_2_Volume thresholds, orange diamonds denote WB nearest-distance settings, and magenta stars denote WB ribbon, default. Yeo macro-averages use 51 and 114 components; Schaefer macro-averages use 100 parcels. Axis ranges differ between atlas families. WB denotes Connectome Workbench.

Table 1. Spatial metrics for Surf_2_Volume and WB mappings against the paired volume reference for Yeo7 (51 spatial components).

| Atlas / setting | Macro-Dice | Macro-Jaccard | Missing reference (%) | Extra-reference (%) | Voxel label accuracy (%) |
|---|---|---|---|---|---|
| Surf_2_Volume 0.025 | 0.590 | 0.435 | 8.56 | 34.39 | 85.29 |
| Surf_2_Volume 0.05 | **0.598** | **0.443** | 12.26 | 31.96 | 82.03 |
| Surf_2_Volume 0.075 | 0.597 | 0.443 | 14.85 | 31.07 | 79.66 |
| WB nearest 1 mm | 0.230 | 0.132 | 84.63 | **23.95** | 14.93 |
| WB nearest 2 mm | 0.416 | 0.271 | 63.06 | 24.33 | 35.89 |
| WB nearest 3 mm | 0.515 | 0.360 | 44.75 | 25.88 | 53.48 |
| WB nearest 5 mm | 0.573 | 0.420 | 19.60 | 32.65 | 76.95 |
| WB nearest 7 mm | 0.558 | 0.405 | **7.13** | 40.41 | **87.60** |
| WB ribbon, default | 0.337 | 0.210 | 73.67 | 24.97 | 25.63 |

Values are descriptive; bold marks the best displayed value in each metric (lower is better for missing and extra-reference fractions). WB denotes Connectome Workbench.

Table 2. Spatial metrics for Surf_2_Volume and WB mappings against the paired volume reference for Yeo17 (114 spatial components).

| Atlas / setting | Macro-Dice | Macro-Jaccard | Missing reference (%) | Extra-reference (%) | Voxel label accuracy (%) |
|---|---|---|---|---|---|
| Surf_2_Volume 0.01 | 0.513 | 0.363 | **5.31** | 46.21 | 84.74 |
| Surf_2_Volume 0.025 | 0.525 | 0.374 | 8.14 | 43.77 | 82.37 |
| Surf_2_Volume 0.05 | **0.533** | **0.381** | 11.80 | 41.65 | 79.33 |
| WB nearest 1 mm | 0.217 | 0.124 | 84.70 | **35.44** | 14.58 |
| WB nearest 2 mm | 0.384 | 0.247 | 63.25 | 35.77 | 35.01 |
| WB nearest 3 mm | 0.469 | 0.320 | 45.03 | 37.08 | 52.11 |
| WB nearest 5 mm | 0.520 | 0.368 | 19.43 | 42.42 | 75.30 |
| WB nearest 7 mm | 0.504 | 0.352 | 7.06 | 49.12 | **85.55** |
| WB ribbon, default | 0.309 | 0.190 | 73.54 | 35.66 | 25.33 |

Values are descriptive; bold marks the best displayed value in each metric (lower is better for missing and extra-reference fractions). WB denotes Connectome Workbench.

Table 3. Spatial metrics for Surf_2_Volume and WB mappings against the paired volume reference for Schaefer7 (100 parcels).

| Atlas / setting | Macro-Dice | Macro-Jaccard | Missing reference (%) | Extra-reference (%) | Voxel label accuracy (%) |
|---|---|---|---|---|---|
| Surf_2_Volume 0.01 | 0.720 | 0.567 | **5.09** | 16.66 | 78.38 |
| Surf_2_Volume 0.025 | 0.728 | 0.577 | 7.95 | 12.89 | 76.27 |
| Surf_2_Volume 0.05 | **0.730** | **0.581** | 11.60 | 9.59 | 73.57 |
| WB nearest 1 mm | 0.235 | 0.135 | 84.93 | **0.65** | 13.33 |
| WB nearest 2 mm | 0.464 | 0.309 | 63.85 | 1.34 | 31.98 |
| WB nearest 3 mm | 0.601 | 0.441 | 45.82 | 3.13 | 47.71 |
| WB nearest 5 mm | 0.709 | 0.556 | 20.57 | 11.22 | 68.94 |
| WB nearest 7 mm | 0.708 | 0.552 | 7.94 | 21.08 | **78.59** |
| WB ribbon, default | 0.358 | 0.226 | 73.83 | 0.86 | 23.30 |

Values are descriptive; bold marks the best displayed value in each metric (lower is better for missing and extra-reference fractions). WB denotes Connectome Workbench.

Table 4. Spatial metrics for Surf_2_Volume and WB mappings against the paired volume reference for Schaefer17 (100 parcels).

| Atlas / setting | Macro-Dice | Macro-Jaccard | Missing reference (%) | Extra-reference (%) | Voxel label accuracy (%) |
|---|---|---|---|---|---|
| Surf_2_Volume 0.01 | 0.719 | 0.566 | **5.09** | 16.66 | 78.35 |
| Surf_2_Volume 0.025 | 0.727 | 0.577 | 7.95 | 12.89 | 76.24 |
| Surf_2_Volume 0.05 | **0.729** | **0.580** | 11.60 | 9.59 | 73.54 |
| WB nearest 1 mm | 0.235 | 0.135 | 84.93 | **0.65** | 13.33 |
| WB nearest 2 mm | 0.464 | 0.309 | 63.85 | 1.34 | 31.98 |
| WB nearest 3 mm | 0.601 | 0.441 | 45.82 | 3.13 | 47.71 |
| WB nearest 5 mm | 0.709 | 0.556 | 20.57 | 11.22 | 68.94 |
| WB nearest 7 mm | 0.708 | 0.552 | 7.94 | 21.08 | **78.59** |
| WB ribbon, default | 0.358 | 0.226 | 73.83 | 0.86 | 23.30 |

Values are descriptive; bold marks the best displayed value in each metric (lower is better for missing and extra-reference fractions). WB denotes Connectome Workbench.

### 3.4 HCP parcel time-series agreement

For Yeo7, participant-level mean CCCs were 0.844 (SD 0.013), 0.849 (SD 0.013), and 0.850 (SD 0.013) at selected thresholds 0.025, 0.05, and 0.075, respectively. WB nearest 5 mm again gave the highest WB CCC (0.821), compared with 0.806 for nearest 7 mm, 0.805 for nearest 3 mm, 0.724 for ribbon, and 0.672 for nearest 1 mm. For Yeo17, thresholds 0.01, 0.025, and 0.05 yielded means of 0.803 (SD 0.016), 0.810 (SD 0.016), and 0.815 (SD 0.016), respectively. The corresponding WB means were 0.793 for nearest 5 mm, 0.781 for nearest 7 mm, 0.772 for nearest 3 mm, 0.688 for ribbon, and 0.638 for nearest 1 mm (Fig. 4B). All selected Yeo thresholds therefore exceeded the strongest WB comparator in time-series agreement.

Against the paired volume references, Schaefer7 and Schaefer17 mean CCCs were 0.939 and 0.939 at threshold 0.01, 0.943 and 0.942 at 0.025, and 0.944 and 0.943 at 0.05, respectively. At 0.05, between-participant SDs were 0.009 for both atlases. WB nearest 5 mm had the highest WB mean CCC (0.927), followed by nearest 7 mm (0.921), nearest 3 mm (0.889), ribbon (0.784), and nearest 1 mm (0.719). Thus, all selected Schaefer thresholds exceeded all tested WB mappings in mean time-series concordance (Fig. 4C).

All 15 planned paired comparisons per atlas favored Surf_2_Volume and remained significant after separate within-atlas FDR correction (all adjusted $p < 0.001$; 60 comparisons across four configurations). At threshold 0.05, the mean CCC improvement over WB nearest 5 mm was 0.028 (95% CI 0.026–0.030) for Yeo7, 0.022 (0.019–0.024) for Yeo17, 0.017 (0.015–0.019) for Schaefer7, and 0.017 (0.015–0.018) for Schaefer17. Even Yeo17 threshold 0.01, which did not exceed WB nearest 5 mm in spatial macro-overlap, had a positive CCC difference of 0.010 (95% CI 0.008–0.012). These results show a consistent temporal-concordance advantage across every selected setting in each tested atlas.

The threshold with the highest temporal concordance was atlas dependent. Yeo7 had its highest selected mean CCC at 0.075, despite attaining its highest spatial overlap at 0.05. Yeo17 and both Schaefer configurations had their highest selected mean CCC at 0.05. Automatic selection therefore retained candidates with different spatial and temporal trade-offs. Agreement was lower in Yeo than in Schaefer, but positive within-atlas differences relative to WB were reproduced across both families.

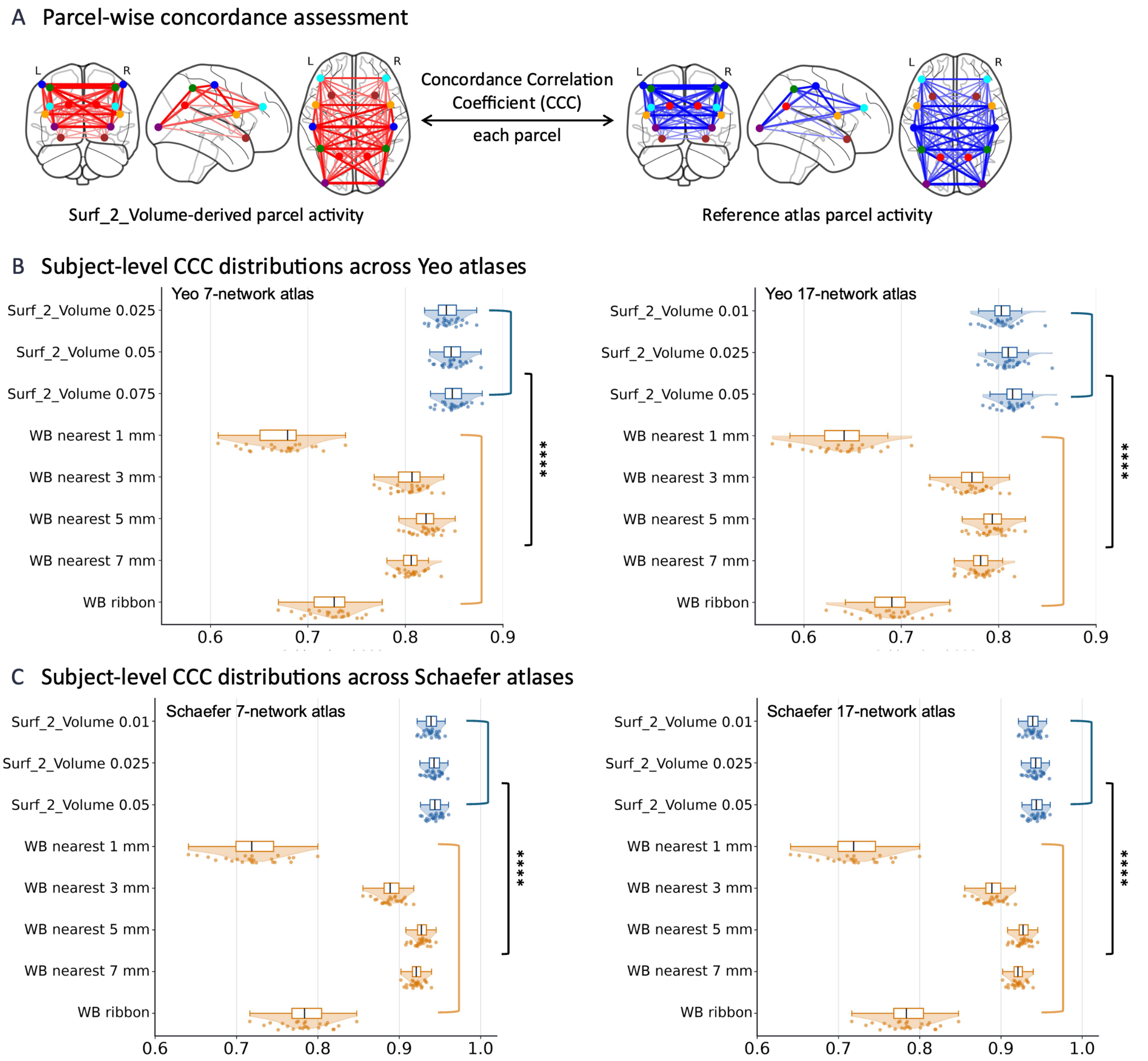

**Fig. 4** Parcel time-series concordance in 24 HCP participants. A: schematic comparison of corresponding parcel signals from a converted atlas and its paired volume reference; inference uses the extracted time series, not the illustrated interregional connections. B: participant-level mean CCC distributions for Yeo7 (left) and Yeo17 (right). C: corresponding distributions for Schaefer7 (left) and 17 (right). Each dot is one participant's mean across the atlas's parcels or components. All panels show three automatically selected Surf_2_Volume thresholds and five WB mappings. Brackets group methods; each package threshold was compared separately with each WB mapping. Asterisks indicate significant comparisons; all 15 tests per atlas had within-atlas FDR-adjusted $p < 0.001$. WB denotes Connectome Workbench.

# 4. Discussion

Surf_2_Volume addresses a practical problem in cortical atlas conversion: how to choose mapping parameters when no published volume counterpart is available. Its contribution combines a projection procedure controlled by the target cortical ribbon with automatic scoring based on source atlas label proportions. The same geometric inputs are used across candidate thresholds, and the source and output label distributions provide an explicit basis for ranking them. This turns parameter exploration into a reproducible workflow that generates, scores, and retains candidate volumes rather than requiring users to judge each setting from appearance alone.

The Yeo and Schaefer results show that this selection criterion identified useful candidates across two atlas families. All four selected sets included the threshold with the highest reference overlap; 11 of the 12 selected outputs exceeded the strongest tested WB macro-Dice, and all 12 exceeded the strongest WB mean time-series CCC. In Yeo7 and Yeo17, threshold 0.05 improved macro-Dice over WB nearest 5 mm by 0.025 and 0.013, respectively, while also reducing missing and extra-reference fractions. The corresponding improvements were 0.021 and 0.020 for Schaefer7 and Schaefer17. Because the paired volume references and HCP data did not contribute to ranking, these findings evaluate the candidates after selection and support the use of source atlas information to guide conversion settings.

The coverage patterns also explain the value of controlling mapping support. Increasing WB nearest distance expands the region allowed to inherit a surface label. In the tested atlases, this reduced missing support but increased labeling outside the paired reference; WB nearest 5 mm had higher overlap and mean CCC than nearest 7 mm despite covering fewer reference voxels. Surf_2_Volume instead changes the eligible region through the smoothed target ribbon and fills that region using categorical votes. At threshold 0.05 it achieved higher macro-Dice and macro-Jaccard than every tested WB mapping, and improved all reported spatial measures relative to WB nearest 5 mm. WB nearest 7 mm still had lower missing support and higher voxel label accuracy, making the choice of evaluation measure relevant.

Returning several high-ranked candidates makes these choices available to users. Yeo7 favored threshold 0.05 for spatial overlap, while 0.075 had a slightly higher mean time-series CCC. Keeping both allows the intended analysis to guide the final choice. The criterion compares relative label proportions rather than local boundaries, so candidate ranking and reference-based overlap need not coincide. The current implementation retains three candidates; the methodological contribution is the scoring and ranking procedure, rather than the particular number of outputs.

The paired references provide concrete targets for evaluating atlas reproduction. Extra-reference labeling denotes assignment in voxels marked as background by the reference, and does not by itself identify their tissue type. Likewise, higher CCC indicates closer agreement with reference-derived parcel signals, including their level and scale. These measures support the tested volume atlas and signal-extraction use cases, while biological boundary accuracy and downstream connectivity or prediction would require separate evaluation.

The Python interface and bundled MNI152NLin6Asym assets reduce preparation for that target space. Other standard spaces remain possible when compatible surfaces, registrations, M2M files, and ribbon references are supplied; their mapping performance was not evaluated here. For analyses in which cortical topology or precise areal localization is central, retaining the original surface parcellation remains important (Coalson et al., 2018).

# 5. Limitations

Validation was limited to four cortical configurations from two atlas families and 24 HCP participants. The two Schaefer configurations share parcel boundaries, and performance in other atlases, target spaces, resolutions, and subcortical labels remains to be tested. The proportion-based score identifies candidates but does not guarantee the best local boundary or signal agreement in a new dataset. Further evaluation should establish how well it transfers to atlases without a published volume counterpart and to additional functional datasets.

# 6. Conclusion

Surf_2_Volume converts CIFTI cortical labels to NIfTI volumes with configurable cortical ribbon support and automatic threshold selection. It scores candidates using the relative representation of source cortical labels and output voxels, without requiring a published volume atlas for selection. This provides a reproducible way to generate and compare volume candidates from a surface parcellation.

Across Yeo7, Yeo17, Schaefer7, and Schaefer17, the automatically selected sets included the highest-overlap tested threshold. At 0.05, spatial macro-overlap exceeded every tested WB mapping. In 24 HCP participants, all three selected outputs per configuration achieved higher mean parcel time-series CCC than all five WB comparators. The results support the approach across two atlas families while showing that coverage and the preferred threshold depend on the evaluation criterion.

The Python interface and bundled target references make this workflow accessible for cortical atlases, including those distributed only in surface space. Retaining several ranked candidates allows users to choose a conversion for their intended analysis, with paired volume references enabling further validation when available.

## Statements and Declarations

### Data availability

The Yeo2011 atlases are available from their cited source (Thomas Yeo et al., 2011); we used the CBIG N1000 SplitLabels representations with 51 and 114 spatial components for the 7- and 17-network configurations. Schaefer2018 atlases are available from their cited source (Schaefer et al., 2018). HCP-Young Adult 2025 data were accessed under the HCP data-use policy and are not redistributed here. The package source is available at https://github.com/thedarkkinght/surf_2_volume.

### Ethics

This secondary analysis used existing de-identified HCP data and involved no new recruitment or data collection. Participant consent and institutional review for the original HCP data collection were handled by the HCP.

### Author contributions

S.Y. and Z.W. (Ziyi Wang) contributed equally and led the methodology, software implementation, data analysis, visualization, and manuscript drafting. Y.S., Z.W. (Zhengye Wang), J.L., and Y.N. contributed to validation, interpretation, and manuscript revision. F.C. and S.W. supervised the study, contributed to study design and interpretation, and critically revised the manuscript.

### Competing interests

The authors declare no competing interests.

### Funding

This work was supported by the National Natural Science Foundation of China (32171051), the Guangdong Key Research and Development Program (2023B0303010004), and the Research Center for Brain Cognition and Human Development, Guangdong (2024B0303390003).

### Software availability

Surf_2_Volume is available from TestPyPI and can be installed with *pip install -i https://test.pypi.org/simple/ surf2volume==0.2.0*.

### AI-use statement

Generative AI assistance was used to draft and translate manuscript text and to edit language. The authors reviewed and are responsible for the scientific content, data, citations, and final wording.